\documentclass[showpacs,floatfix,aps,prl,preprint,superscriptaddress]{revtex4-1}

\usepackage{amsfonts} 
\usepackage{amsmath}
\usepackage{amssymb}
\usepackage{braket}
\usepackage{breqn}
\usepackage{bm} 
\usepackage{upgreek}

\usepackage{dcolumn}
\usepackage{etoolbox}

\usepackage{float}
\usepackage{graphicx}
\usepackage{lipsum}
\usepackage{mathtools}

\usepackage{verbatim}
\usepackage[utf8]{inputenc} 
\usepackage{soul,xcolor,colortbl}
\usepackage{hyperref}\hypersetup{colorlinks=true,citecolor=blue,linkcolor=blue,urlcolor=blue}

\begin{document}

\title{Phase Pattern Formation in Grain Boundaries}
\author{I. S. Winter}\email{iswinte@sandia.gov}
\affiliation{Sandia National Laboratories$,$ Albuquerque$,$ NM$,$ 87123$,$ USA}

\author{T. Frolov}
\email{frolov2@llnl.gov}
\affiliation{Lawrence Livermore National Laboratory$,$ Livermore$,$ CA$,$ 94550$,$ USA}
\date{\today}
\begin{abstract}
In this work we derive conditions that predict the existence of two-phase periodic-pattern grain boundary structures that are stable against coarsening. While previous research has established that elastic effects can lead to phase pattern formation on crystal surfaces, the possibility of stable grain boundary structures composed of alternating grain boundary phases has not been previously analyzed. Our theory identifies the specific combination of grain boundary and materials properties that enable the emergence of patterned grain boundary states and shows that the dislocation content of grain boundary phase junctions, absent in surface phenomena, weakens the stability of the patterned structures. The predictions of the theory are tested using a model copper grain boundary that exhibits multiple phases and two-phase pattern formation. We discuss how, similarly to surfaces, elastic effects associated with grain boundary phase junctions have profound implications for how grain boundary phases transform.

\end{abstract}
\maketitle

Elastic effects arising from interactions between junctions of surface phases are known to have a profound effect on phase transformations in 2D \cite{marchenko1981possible,Hannon2001}. Line forces, also known as force monopoles, exist at surface phase junctions due to the difference in surface stress \cite{marchenko1980elastic,Alerhand1988,Pehlke1991,marchenko1981possible,SROLOVITZ1991111}. A seminal study by Marchenko showed that the interaction of these surface line defects can in some cases reduce the surface energy of the system relative to a coarse two-phase state. This stabilizes a new surface structure, which can be described as a periodic pattern composed of alternating surface phases \cite{marchenko1980elastic}. The characteristic dimensions of the pattern depend on the temperature, composition, and elastic properties of the system. 

Conventional thermodynamics, which does not consider long-range interactions, predicts surface phase diagrams with well-defined transition boundaries that correspond to the equilibrium coexistence of the different phases. In reality, elastic interactions can smear the coexistence lines defined by the equality of free energies, creating a new domain represented by the patterned state. Patterned surface states with a well-defined pattern wavelength and no well-defined transition temperature have been experimentally observed and confirmed for surface phases of silicon \cite{Hannon2001}.

Grain boundaries are internal interfaces inside a material that, just like surfaces, exhibit different phases \cite{CANTWELL20141,krause2019review,cantwell2020grain}. However,  the possibility of metastable pattern formation at GBs is currently not known. Just as for surfaces, the junctions between different GB phases give rise to line forces, because different GB phases have different GB stresses \cite{WinterGBphases2022}. In this regard, GB phases are equivalent to surface phases and have the potential for forming patterned states. However, one crucial distinction is that GB phase junctions also contain dislocations \cite{FrolovBurgers,WinterGBphases2022,WINTER2022118067}. This additional dislocation-dislocation interaction energy contribution can counteract the elastic energy of line forces, prohibiting the formation of patterned GB states. In this work, we derive equations describing the energy of the patterned states at GBs that include both line forces and dislocation interactions and demonstrate conditions under which patterned GB states can exist. Using the predictions of our model we identify a copper model GB system that exhibits pattern behavior. Our simulations demonstrate that the structure composed of alternating GB phases is stable against coarsening at finite temperatures.

\begin{figure}[ht!]
\centering
\includegraphics[width=\columnwidth]{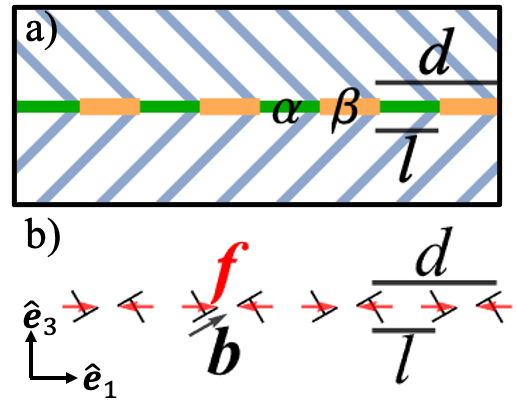}
\caption{Panel a depicts a schematic of a periodic GB phase pattern, with the $\alpha$ and $\beta$  GB phases represented by green and orange regions. The repeat length of the pattern is $d$ and the length of the $\alpha$ phase is $l$. Panel b depicts the corresponding periodic pattern of GBPJs represented as line forces and dislocations.}
\label{fig:schematic_pattern}
\end{figure}

In a manner analagous to Hamilton \textit{et al.}'s study of GB facets \cite{Hamilton2003}, let us consider a periodic GB phase pattern consisting of two GB phases as depicted in Fig. \ref{fig:schematic_pattern}. We will term the two GB phases to be the $\alpha$ and $\beta$ phases. The repeat distance of the pattern is of length $d$, and the lengths of the $\alpha$ and $\beta$ phases are $l$ and $l-d$ respectively. Both a line force and dislocation can exist at the GB phase junction \cite{WinterGBphases2022}. The line force arises due to the difference in GB stresses: $\bm{f} = \left(\tau_{11}^{\alpha} - \tau_{11}^{\beta}\right)\hat{\bm{e}}_1$. The dislocation is a result of the difference in excess volume, excess shear, and the density of atoms at the GB for the two GB phases \cite{FrolovBurgers}. The energy per unit length of such a two-phase pattern depicted in Fig. \ref{fig:schematic_pattern} is expressed in terms of $\chi=l/d$ as

\begin{equation}\label{eq:pattern1_nucleation}
    E(\chi) = \Delta \gamma^{\alpha\beta}\chi+ \frac{A}{d}\ln\left( \frac{d \sin(\pi \chi)}{\pi\rho} \right) + \frac{2\Gamma^{\alpha\beta}+B}{d},
\end{equation}

\noindent where $\Delta \gamma^{\alpha\beta} = \gamma^{\alpha}-\gamma^{\beta}$ is the difference in GB energy between the $\alpha$ and $\beta$ GB phases. $A$ and $B$ are elastic terms related to $\bm{f}$, $\bm{b}$, and the bulk elastic constants of the material. $\rho$ and $\Gamma^{\alpha\beta}$ are the core radius and core energy per unit-length of the GBPJ. A derivation of Eq. \eqref{eq:pattern1_nucleation} is given in the Supplemental Materials.

From Eq. \eqref{eq:pattern1_nucleation} it can be shown that a local maximum or minimum energy exists for the dual-phase system if $\frac{dE(\chi)}{d\chi}=0, \chi\in(0,1)$. For pattern formation to be at least metastable, this extreme point, $\chi_c$, must be a local minimum:$ \frac{d^2E(\chi)}{d\chi^2}|_{\chi=\chi_c}>0$. As $\frac{d^2E(\chi)}{d\chi^2} = -\frac{\pi^2 A}{d \sin^2(\pi\chi)}$, the possibility for pattern formation to occur is entirely dependent on the sign of the elastic coefficient, $A$. If $A < 0$, a pattern can exist in a metastable state.

To better understand the conditions necessary for pattern formation to occur, we consider the case of an elastically isotropic system, as this allows us to write an analytical expression for $A$ \cite{WinterGBphases2022}:

\begin{subequations}\label{eq:defineA}
    \begin{align}
        A &= A^{dd}+A^{pp}+A^{dp},\\
         A^{dd} &= \mu \Big[(1-\nu)b_2^2 + (b_1^2+b_3^2)^2\Big]/2\pi(1-\nu),\\
         A^{pp} &= -f_1^2 (3-4\nu)/8\pi\mu(1-\nu),\\
         A^{dp} &= -f_1 b_3 (1-2\nu)/2\pi(1-\nu),
    \end{align}
\end{subequations}


\noindent where $\mu$ is the shear modulus, $\nu$ Poisson's ratio and $b_i$ the components of the Burgers vector. The first term in $A$, the dislocation-dislocation interaction energy, must be positive if the material is to be elastically stable. The second term, the line force-line force interaction must be negative if the material is to be elastically stable. The line force-dislocation interaction can be positive or negative depending on the signs of $f_1$ and $b_3$. 

Let us now try to find what values of $f_1$ allow for $A<0$. $A$ is a quadratic equation with respect to $f_1$, so finding the roots of $f_1$ is instructive:
\begin{widetext}
\begin{subequations}\label{eq:roots}
\begin{align}
    f^{0,low}_1 &= -\frac{2\mu}{3-4\nu}\left(b_3(1-2\nu) + \sqrt{b_1^2(3-4\nu) + (1-\nu)\Big[ 4b_3^2(1-\nu) + b_2^2(3-4\nu)\Big]}\right),\\
    f^{0,high}_1 &= -\frac{2\mu}{3-4\nu}\left(b_3(1-2\nu) - \sqrt{b_1^2(3-4\nu) + (1-\nu)\Big[ 4b_3^2(1-\nu) + b_2^2(3-4\nu)\Big]}\right).
\end{align}
\end{subequations}  
\end{widetext}

\noindent Mechanical stability  requires that $-1<\nu<1/2$, so the roots of $f_1$ must be real for an elastically stable material. A necessary condition for pattern formation, $A < 0$, is met if $f_1 > f^{0,high}_1$ or $f_1 < f^{0,low}$. Thus the theory allows us to formulate the following screening criteria for GBs capable of pattern formation.  GB phases should have large differences in GB stresses, a small Burgers vector at the GBPJ, and a small shear modulus of the material.




Following the criteria identified by the theory we consider the possibility of pattern formation in a face-centered-cubic Cu $\Sigma 5 (210)[001]$ GB composed of repeating units of split-kite and filled-kite GB phases \cite{frolov_structural_2013}. These two GB phases have a large difference in $\tau_{11}$ as shown in TABLE \ref{tab:phase_info} resulting in a large line force, $f_1 = -4.66$ J$/$m$^2$, while having a relatively small Burgers vector, $b<0.5\ \mathrm{\AA}$ \cite{FrolovBurgers}. To match the boundary dimensions along the tilt axis in our coexistence simulations we selected one of the metastable, but essentially energy degenerate compared to the ground state, split-kite structure \cite{Zhu2018}.

\begin{table}[hbt!]
\caption{\label{tab:phase_info} Properties of the two grain boundary phases considered in this work.}
\begin{ruledtabular}
\begin{tabular}{llllll}
GB phase & Split kite & Filled kite\\
GB Energy (J$/$m$^2$) & 0.942 & 0.953\\
$\tau_{11}$ (J$/$m$^2$) & -2.36 & 2.30\\
Structural Unit Area ($a_0^2$) &  $\sqrt{5}\times2$ & $7\sqrt{5}\times2$ \\
Fraction of Plane & 1/2 & 6/7
\end{tabular}
\end{ruledtabular}
\end{table}


Following the work of Barnett and Lothe \cite{Barnett_1974}, we applied anisotropic linear elasticity theory to calculate the elastic coefficient in Eq. \eqref{eq:pattern1_nucleation}, $A$, using an embedded-atom method potential of Cu \cite{PhysRevB.63.224106} within the molecular dynamics simulation code, LAMMPS \cite{LAMMPS}. Eq. \eqref{eq:pattern1_nucleation} remains valid in the anisotropic case, but in general the elastic coefficient $A$ does not necessarily have an analytical form. At 0 K, $A$ was calculated from the elastic constants of the system, $\tau_{11}$ for each GB phase (given in TABLE \ref{tab:phase_info}), and the Burgers vector of the GBPJ. Using the approach outlined in Ref. \cite{FrolovBurgers}, we found the Burgers vector of the GBPJ to be $\bm{b} = 0.368\hat{\bm{e}}_3$. From these inputs we find $A=-31$ meV$/$\AA, and as a result, predict pattern formation to occur.

\begin{figure}[ht!]
\centering
\includegraphics[width=\columnwidth]{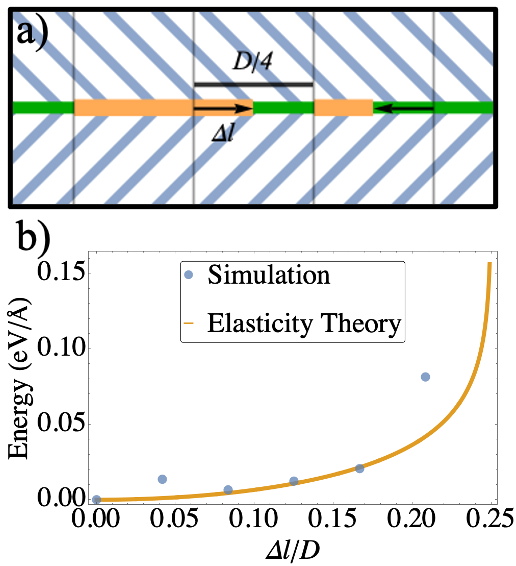}
\caption{Panel a depicts the double-period GB-phase-patterned state. Panel b shows a comparison of the energy of a double-period structure using molecular statics simulations and elasticity theory. The increasing energy shows that the patterned state is stable against coarsening.}
\label{fig:ms_pattern}
\end{figure}


To test if the magnitude of $A$ is large enough to maintain pattern stability we first consider 0 K molecular statics simulations of the effect of pattern morphology on system energy. Since the two different grain boundary phases are composed of a different number of atoms (TABLE \ref{tab:phase_info})  and the total number of atoms in the simulation box is fixed, a heterogeneous state composed of just two phases is always stable in a wide range of temperatures. As a result, to investigate the stability against coarsening, we have to consider a double-period system consisting of four GB phases repeating over a length $D$, two split kites and two filled kites as shown in Fig. \ref{fig:ms_pattern}a. The double-period system is constructed such that the total length of split- and filled-kite islands is constant, and as a result the number of atoms in the system too: if one split-kite island grows in length, the other shrinks. If we constrain the larger (smaller) split-kite island to have the same length as the larger (smaller) filled-kite island, we can describe the system by one parameter, $\Delta l$: the difference in length between the two islands of a given GB phase junction, shown in Fig. \ref{fig:ms_pattern}. The elastic energy from such a double-period structure can be written as

\begin{equation}\label{eq:double-period}
    \Delta E^{DP}\left(\Delta l\right) = A\ln \left( \cos\left(\frac{2\pi \Delta l}{D}\right) \right).
\end{equation}

\noindent The derivation of Eq. \eqref{eq:double-period} is detailed in the Supplemental Materials. In Fig. \ref{fig:ms_pattern}, Eq. \eqref{eq:double-period} is compared to molecular statics simulations of the double-period system as a function of $\Delta l$, which are generated using the method described in Ref. \cite{WinterGBphases2022}. As the simulations contained approximately $500,000$ atoms, the molecular statics simulations of the double-period structure were performed by annealing the structures at 300 K for 20 ns and then relaxing the structure using the FIRE optimization algorithm \cite{FIRE}. While the grain boundary structures were already in their lowest energy states from the beginning, the structure of the GBPJ cores required additional optimization which occurred during the annealing stage. The GB energy of all systems converged to $10^{-8}$ eV$/\mathrm{\AA}^2$. However, due to the large system size and complex GB microstructure, the scatter of the 0 K energies is still seen in Fig. \ref{fig:ms_pattern}. 
Overall, however, we see an excellent agreement between the analytical prediction of equation \eqref{eq:double-period} and direct GB energy calculations of the patterned state using molecular statics at 0K. Both calculations suggest that the regular patterned state should be stable against coarsening.



\begin{figure}[ht!]
\centering
\includegraphics[width=\columnwidth]{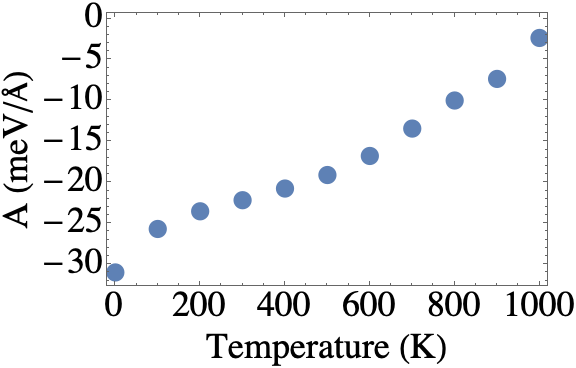}
\caption{The elastic coefficient, $A$, plotted as a function of temperature confirms the stability of the patterned state at finite temperatures. Its decreasing magnitude suggests that the temperature decreases the stability of the patterned state. Error bars are smaller than the size of the data points.}
\label{fig:prefactor_temp}
\end{figure}

To investigate the effect of temperature on the stability of the patterned state, we calculated $A$ in the temperature interval from 0 to 1000 K. This required the calculation of the elastic constants, $\tau_{11}$, and thermal expansion as a function of temperature, the details of which are given in the Supplemental Materials.
The calculated elastic coefficient $A(T)$ is shown in Fig. \ref{fig:prefactor_temp}. It remains negative, predicting the stability of the patterned state at finite temperature. The temperature however has a destabilizing effect as $A$ gradually decreases in magnitude and reaches a near zero value at 1000 K, marking the limit of the patterned state's stability.


\begin{figure}[ht!]
\centering
\includegraphics[width=\columnwidth]{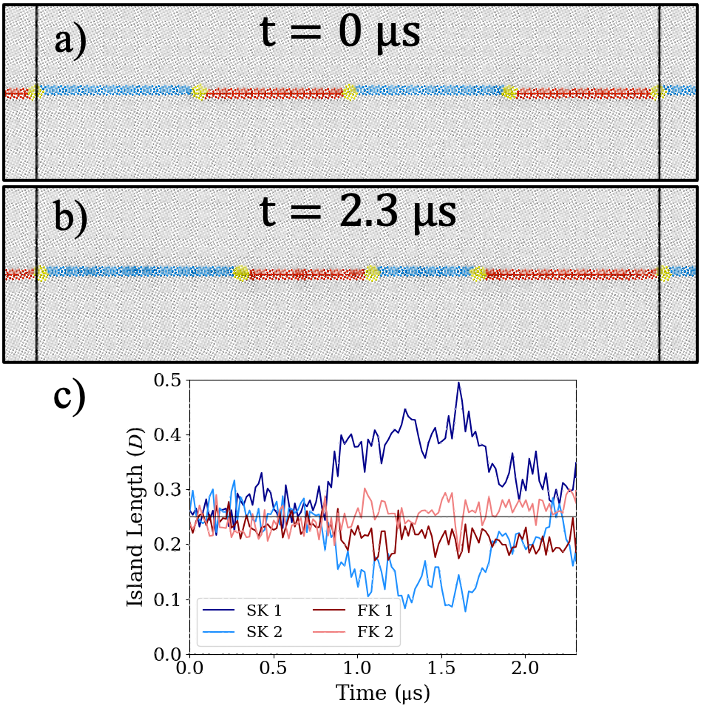}
\caption{Panels a and b depict the initial and final island configurations. Blue atoms correspond to split-kites islands, red atoms correspond to filled-kite islands, and yellow atoms correspond to GB phase junctions. Panel c shows the evolution of islands over time. }
\label{fig:md_islands}
\end{figure}

Finally, we tested the predictions of our theory and calculations by directly simulating the evolution of the double-period patterned state at high temperature using molecular dynamics in the NVT ensemble. The patterned state containing four GB phase islands shown in Fig. \ref{fig:md_islands} should be stable against coarsening into a two-phase state. For our MD simulations, we selected 800 K as the optimal testing temperature. At higher temperatures, the stabilizing force becomes vanishingly small and comparable to fluctuations, so we expect the pattern to go away. At lower temperatures, the island size fluctuations become too infrequent to provide convincing proof of stability. The initial patterned state with all four islands of equal length, $l=D/4$, is shown in Fig. \ref{fig:md_islands}a. We ran the simulation for a total of 2.3 $\upmu$s. 

We estimated the lengths of the four islands, shown in Fig. \ref{fig:md_islands}c, by calculating the contour of the GB (and the step heights between GB phase junctions). The contour was estimated by applying polyhedral template matching \cite{Larsen_2016}, as implemented in OVITO \cite{ovito}, with non-FCC atoms considered to be part of the GB. Hyperbolic tangent functions were fit to the GB contour to calculate island lengths, with more details of the simulation given in the Supplemental Materials. The GB pattern showed remarkable stability during the first 0.75 $\upmu$s, during which the length of all islands fluctuated around the equal size state with $l=D/4$, as shown in Fig. \ref{fig:md_islands}c. We note that states with different $l$ are visited during this time interval as the junctions migrate by several GB units (see movie in the Supplemental Materials). 

At approximately t=0.75 $\upmu$s we observe a large fluctuation towards a coarser state when the length of one of the split-kite islands increases significantly at the expense of the other split-kite island. However, at approximately 1.75 $\upmu$s, the system returns to the patterned state, confirming its stability. The state of the system after 2.3 $\upmu$s is shown in Fig. \ref{fig:md_islands}b. 

The observed large fluctuation of the island size away from the even size state at 800 K is indicative of an already very small stabilizing contribution from the elastic interactions compared to thermal fluctuations. We expect that at higher temperatures thermal fluctuations will dominate the evolution of such a microstructure eventually leading to coarsening. At lower temperatures on the other hand we expect the patterned state to be stable.



There is an increasing recognition that just like in 3D materials, complex defect microstructures exist inside interfaces and greatly influence the properties of polycrystalline materials. GBs may contain ordered or random networks of disconnections and/or GB phase junctions. These defected GB microstructures could form during non-equilibrium processes such as plastic deformation, grain growth, exposure to fluxes of point defects during radiation damage, and other processes that involve changes in temperature and chemical composition.

Our understanding of the dynamic evolution and stability of these GB defect microstructures is currently severely limited by the lack of techniques that enable the identification and visualization of these defects in the GB plane both experimentally and in simulations. Only recently algorithms capable of robust identification of disconnection and GB phase junctions have been proposed \cite{WINTER2022118067,DEKA2023119096}. Despite these limitations, there is growing interest in understanding the collective behavior of disconnections and GBPJs. For example, recent theoretical studies predicted GB topological transitions due to disconnections \cite{chen_grain-boundary_2020}, which could result in dramatic changes in GB mobility \cite{olmsted_grain_2007}. 

Complex dual-phase patterns composed of different GB phases have recently been investigated by high-resolution transmission electron microscopy (HRTEM) in Cu GBs \cite{Meiners,langenohl_dual_2022}, with dual-phase patterned GB microstructures having been directly observed by HRTEM in $\langle111\rangle$ tilt GBs in Cu \cite{langenohl_dual_2022}. To explain the pattern metastability the authors of the study postulated that an array of pre-existing disconnections inherited from processing could exist at the boundary and these disconnections would effectively repel GB phase junctions and make this dual-phase pattern structure metastable. Using our analysis, we can estimate the elastic coefficient to be $A=40$ meV$/$\AA, which means that this particular Cu GB (without additional pre-existing disconnections) should not form a metastable patterned state. We conclude that the observed state is either a result of an incomplete transformation or indeed the presence of a preexisting array of additional disconnections.


Inspired by well-known theoretical and experimental results for surface phases, here we investigated the stability of patterned states of GBs composed of different GB phases separated by line defects containing both line forces and dislocation defects. We derived equations describing the energy of such states and identified conditions where the pattern state becomes metastable, providing a barrier to coarsening into a more common two-phase structure. We show that the Burgers content present at GB phase junctions has a destabilizing effect on the pattern state. Using our theoretical model we identified a model system that should exhibit pattern behavior and verified it by direct atomistic simulations at finite temperature as well as energy calculations at 0 K.


\begin{acknowledgements}
This work was performed under the auspices of the U.S. Department of Energy (DOE) by Lawrence Livermore National Laboratory under contract DE-AC52-07NA27344.  TF was supported by the U.S. DOE, Office of Science under an Office of Fusion Energy Sciences Early Career Award. Computing support for this work came from the Lawrence Livermore National Laboratory Institutional Computing Grand Challenge program. Sandia National Laboratories is a multi-mission laboratory managed and operated by National Technology \& Engineering Solutions of Sandia, LLC (NTESS), a wholly owned subsidiary of Honeywell International Inc., for the U.S. Department of Energy’s National Nuclear Security Administration (DOE/NNSA) under contract DE-NA0003525. TF is grateful to Mark Asta for stimulating discussions.
\end{acknowledgements}

\bibliography{aapmsamp}

\end{document}